\documentclass[epj]{webofc}
\usepackage[varg]{txfonts}   % Web of Conferences font
\woctitle{XLVI International Symposium on Multiparticle Dynamics}
\begin{document}
\title{Muon bundles from extensive air showers}

\author{Maciej Rybczy\'{n}ski\inst{1}\fnsep\thanks{\email{maciej.rybczynski@ujk.edu.pl}} \and
        Zbigniew W\l odarczyk\inst{1}\fnsep\thanks{\email{zbigniew.wlodarczyk@ujk.edu.pl}} \and
        Grzegorz Wilk\inst{2}\fnsep\thanks{\email{Grzegorz.Wilk@ncbj.gov.pl}}
        % etc.
}

\institute{Institute of Physics, Jan Kochanowski University, 25-406 Kielce, Poland
\and
National Centre for Nuclear Research, Department of Fundamental Research, 00-681 Warsaw, Poland
          }

\abstract{%
In this talk we present a possible explanation of the presence of high muon multiplicity events registered recenty by CERN ALICE experiment in its dedicated cosmic ray run.
}
\maketitle
\section{Introduction}
\label{intro}
A very important aspect of understanding of primary cosmic ray flux
and its composition is a proper description of recent measurement done
by the ALICE experiment at CERN LHC, in its dedicated cosmic ray run~\cite{ALICE:2015wfa}.
ALICE Collaboration registered the presence of large groups of muons
produced in EAS by cosmic ray interactions in the upper atmosphere~\cite{ALICE:2015wfa}.
A special emphasis has been given to the study of high multiplicity events
containing more than 100 reconstructed muons.
Similar events have been studied in previous underground experiments
such as ALEPH~\cite{Avati:2000mn}, DELPHI~\cite{Abdallah:2007fk} and L3~\cite{Achard:2004ws}
at CERN Large Electron-Positron Collider.
While these experiments were able to reproduce the measured muon
multiplicity distribution with Monte Carlo simulations at low and intermediate
multiplicities, their simulations failed to describe the frequency
of the highest multiplicity events. The muon multiplicity  distribution
measured by ALICE when compared with the fits obtained from CORSIKA simulations
with proton or iron primary cosmic rays indicates that the expected rate
of higher multiplicity muon events is sensitive to assumptions made about
the dominant hadronic production mechanisms in air shower development.
Although, assuming the presence of only the Fe nuclei in the primary flux it
is roughly possible to describe the presence of the low and intermediate
muon bundles, the same assumption  used in the conventional  models of
EAS development completely fails to describe observed muon groups at
high multiplicities. Surprisingly, the ALICE Collaboration shows in~\cite{ALICE:2015wfa} only
fits for muon multiplicities up to about 70, neglecting many events with measured
muon multiplicities up to 270.

In this talk we discuss the hypothesis that muon bundles of extremely high
multiplicity observed recently by ALICE detector can originate from small
lumps of Strange Quark Matter (SQM) colliding with the atmosphere.
We demonstrate that extremely large groups of muons can be very well
described by a relatively minute (of the order of $10^{-5}$ of total primary flux) admixture
of SQM of the same total energy. Our estimate of SQM flux do not
contradict results obtained recently by the SLIM Collaboration~\cite{Sahnoun:2008mr}.

\section{Strange Quark Matter in cosmic rays}
\label{sec-1}
Following~\cite{Wilk:1996me,Wilk:1996jpg} it is fully sensible to search for SQM in cosmic ray experiments because the specific features of strangelets~\cite{Alcock:1985vc} allow them to penetrate deep into atmosphere~\cite{Wilk:1996je,Gladysz:1997jpg,Ryb:2002mww,Ryb:2006mww}. The point is that there exists some critical size of the strangelet given by the critical value of its mass number, $A=A_{crit}\sim 300 - 400$, such that for $A > A_{crit}$ strangelets are absolutely stable against neutron emission. Below this limit strangelets decay rapidly evaporating neutrons.

\begin{figure}[h]
\begin{center}
\includegraphics[width=10cm]{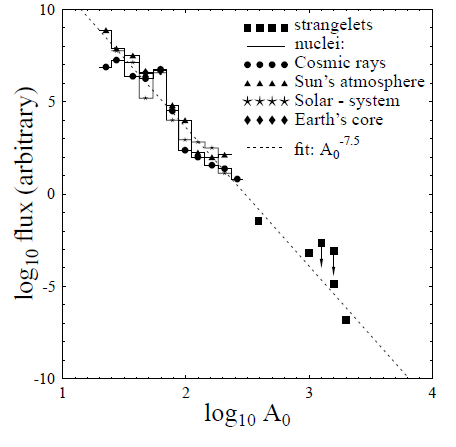}
\caption{\label{fig:comp} Comparison of the estimated mass spectrum $N(A_{0})$ for strangelets with
the known abundance of elements in the Universe~\cite{Rybczynski:2004rr}.}
\end{center}
\end{figure}

The geometrical radii of strangelets turn out to be comparable to the radii of ordinary nuclei~\cite{Wilk:1996me}, i.e., their geometrical cross sections are similar to the normal nuclear ones. To account for their strong penetrability one has to accept that strangelets reaching deeply into atmosphere are formed in many successive interactions with air nuclei by the initially very heavy lumps of SQM entering the atmosphere and decreasing due to the collisions with air nuclei (until their $A$ reaches the critical value $A_{crit}$~\cite{Wilk:1996me}). Such scenario is fully consistent with all present experiments~\cite{Wilk:1996me}. In this scenario interaction of strangelet with target nucleus involves all quarks of the target located in the geometrical intersection of the colliding air nucleus and strangelet.

There are several reports suggesting existence of direct candidates for SQM~\cite{Saito:1990ju} (characterized mainly by their very small ratios of $Z/A$). All of them have mass numbers $A$ near or slightly exceeding $A_{crit}$. Analysis of these candidates for SQM shows~\cite{Wilk:1996me} that the abundance of strangelets in the primary cosmic ray flux is $F_{S}\left(A=A_{crit}\right)/F_{tot}\sim 2.4\cdot 10^{-5}$ at the same energy per particle. For normal flux of primary cosmic rays~\cite{Shibata:1995xz} the expected flux of strangelets is then equal to $F_{S}=7\cdot 10^{-6}~{\rm m}^{-2}{\rm h}^{-1}{\rm sr}^{-1}$ for the energy above 10 GeV per initial strangelet.

The experimental data mentioned before lead to the flux of strangelets which is consistent with the astrophysical limits and with the upper limits given experimentally~\cite{Price:1988ge}. It follows the $A^{-7.5}$ behaviour, which coincides with the behaviour of abundance of normal nuclei in the Universe, see Fig.~\ref{fig:comp} for comparison.

Chart of nuclides presented in Ref.~\cite{Crawford:1994cn} shows all known forms of stable matter. Between the heaviest atomic elements and neutron stars, which are giant nuclei, lies a vast, unpopulated nuclear desert. This void may actually be filled with strange quark matter.

\section{High multiplicity muon bundles from SQM}

\begin{figure}[h]
\vspace{-3mm}
\begin{center}
\includegraphics[width=14cm]{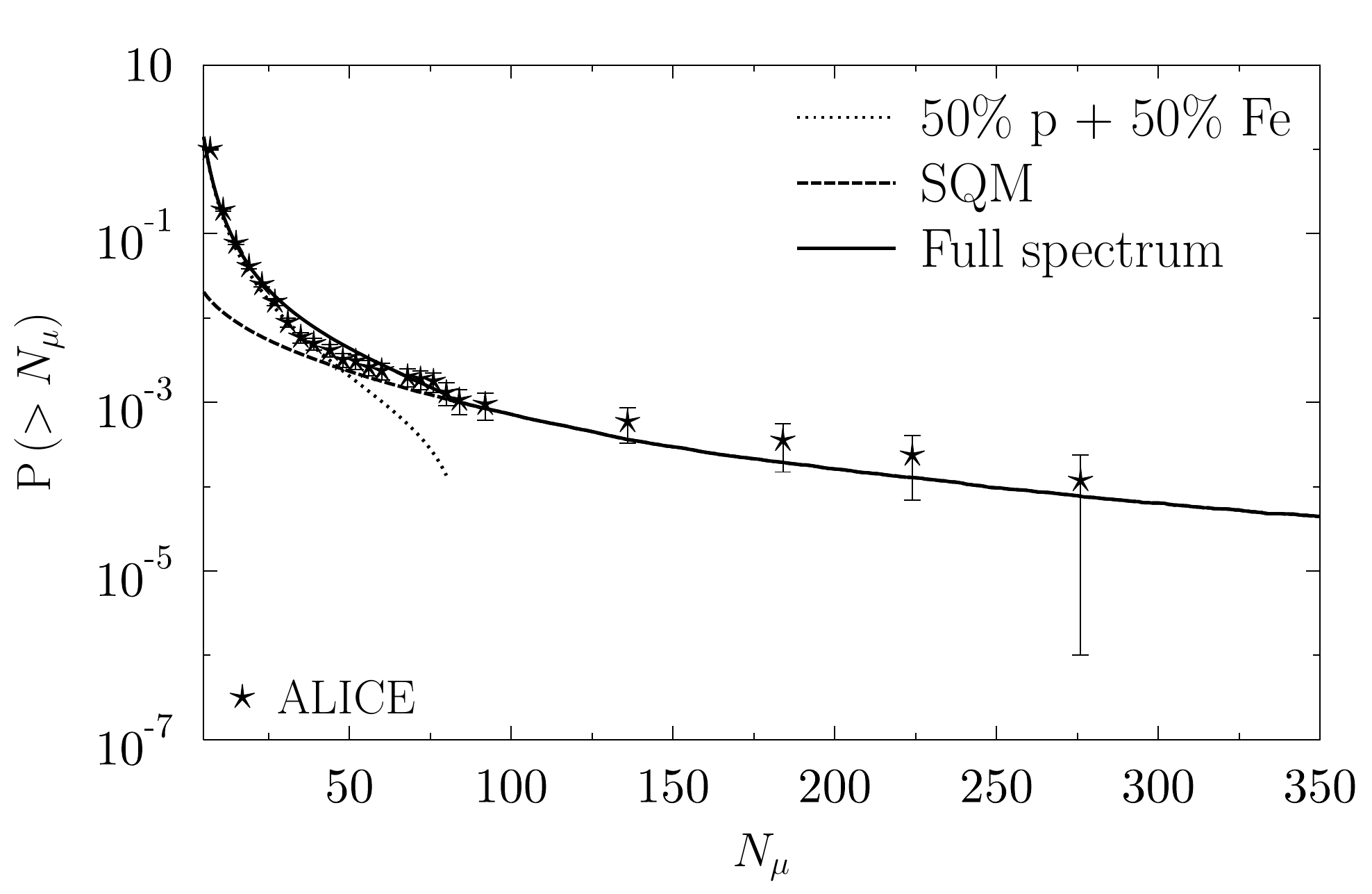}
\caption{Integral  multiplicity  distribution of muons for the ALICE data~\cite{ALICE:2015wfa} (stars). Monte Carlo simulations for primary nuclei composed with 50\% of protons and 50\% of iron nuclei (dotted line) and for primary strangelets with mass A taken from the $A^{-7.5}$ distribution (broken line). Full line shows the summary (calculated) distribution. Simulations were done using suitable modifications of the SHOWERSIM~\cite{SHOWERSIM84} modular software.}
\label{fig:hmm}
\end{center}
\end{figure}

In this study we have used the suitably modified SHOWERSIM~\cite{SHOWERSIM84} modular software. We performed Monte Carlo simulations for primary nuclei composed with 50\% of protons and 50\% of iron nuclei and for primary strangelets with mass A taken from the $A^{-7.5}$ distribution. In Fig.~\ref{fig:hmm} the results of our simulations are shown. The lower and medium multiplicities can be reproduced by the ordinary nuclei. The extremely large groups of muons can be very well described by a relatively minute (of the order of $10^{-5}$ of total primary flux) admixture of SQM of the same total energy.

\section{Conclusions}
\label{sec:concl}

Our conclusions are as follows:

\begin{itemize}
\item Accelerator apparata can be suitable for cosmic-ray physics: LEP experiments were the pioneers on this topic. The LHC ALICE have interesting results, apart from the global physics studies used in model tuning of hadronic interactions.
\item The low multiplicities of muon groups measured by the CERN ALICE experiment favor light nuclei as primaries, whereas medium multiplicities show behavior specific for heavier primaries.
\item At high multiplicities of muon groups the common interaction models fail to describe muon bundles.
\item A relatively small (of the order of $10^{-5}$ of total primary flux) admixture of SQM of the same total energy allows to reproduce the high muon multiplicity groups.
\item Our estimate of SQM flux do not contradict the results obtained recently by the SLIM Collaboration~\cite{Sahnoun:2008mr}.
\end{itemize}

Acknowledgements: Research supported by the Polish National Science Centre grant UMO-2015/18/M/ST2/00125

\end{document}